\documentclass[fleqn,10pt]{wlscirep}
\usepackage[utf8]{inputenc}
\usepackage[T1]{fontenc}
\usepackage{amsmath}
\usepackage{amsfonts}
\usepackage{amssymb}
\usepackage{eurosym}
\usepackage{graphicx}
\usepackage{hyperref}
\usepackage{comment}
\usepackage{verbatim}

\usepackage{authblk}

\title{Superintense Laser-driven Ion Beam Analysis}
\author[1,*]{M. Passoni}
\author[1,**]{L. Fedeli}
\author[1]{F. Mirani}
\affil[1]{\emph{Politecnico di Milano, Department of Energy, Milan, 20133, Italy}}
\affil[*]{matteo.passoni@polimi.it}
\affil[**]{luca.fedeli@polimi.it }
\date{}                     
\begin{abstract}
Ion beam analysis techniques are among the most powerful tools for advanced material characterization. Despite their growing relevance in a widening number of fields, most ion beam analysis facilities still rely on the oldest accelerator technologies, with severe limitations in terms of portability and flexibility. In this work we thoroughly address the potential of superintense laser-driven proton sources for this application. We develop a complete analytical and numerical framework suitable to describe laser-driven ion beam analysis, exemplifying the approach for Proton Induced X-ray/Gamma-ray emission, a technique of widespread interest. This allows us to propose a realistic design for a compact, versatile ion beam analysis facility based on this novel concept. These results can pave the way for ground-breaking developments in the field of hadron-based advanced material characterization.
\end{abstract}

\begin{document}

\flushbottom
\maketitle
%
%
\thispagestyle{empty}

\section*{Introduction}
\begin{figure}[ht]
\centering
\includegraphics[width = 0.9\textwidth]{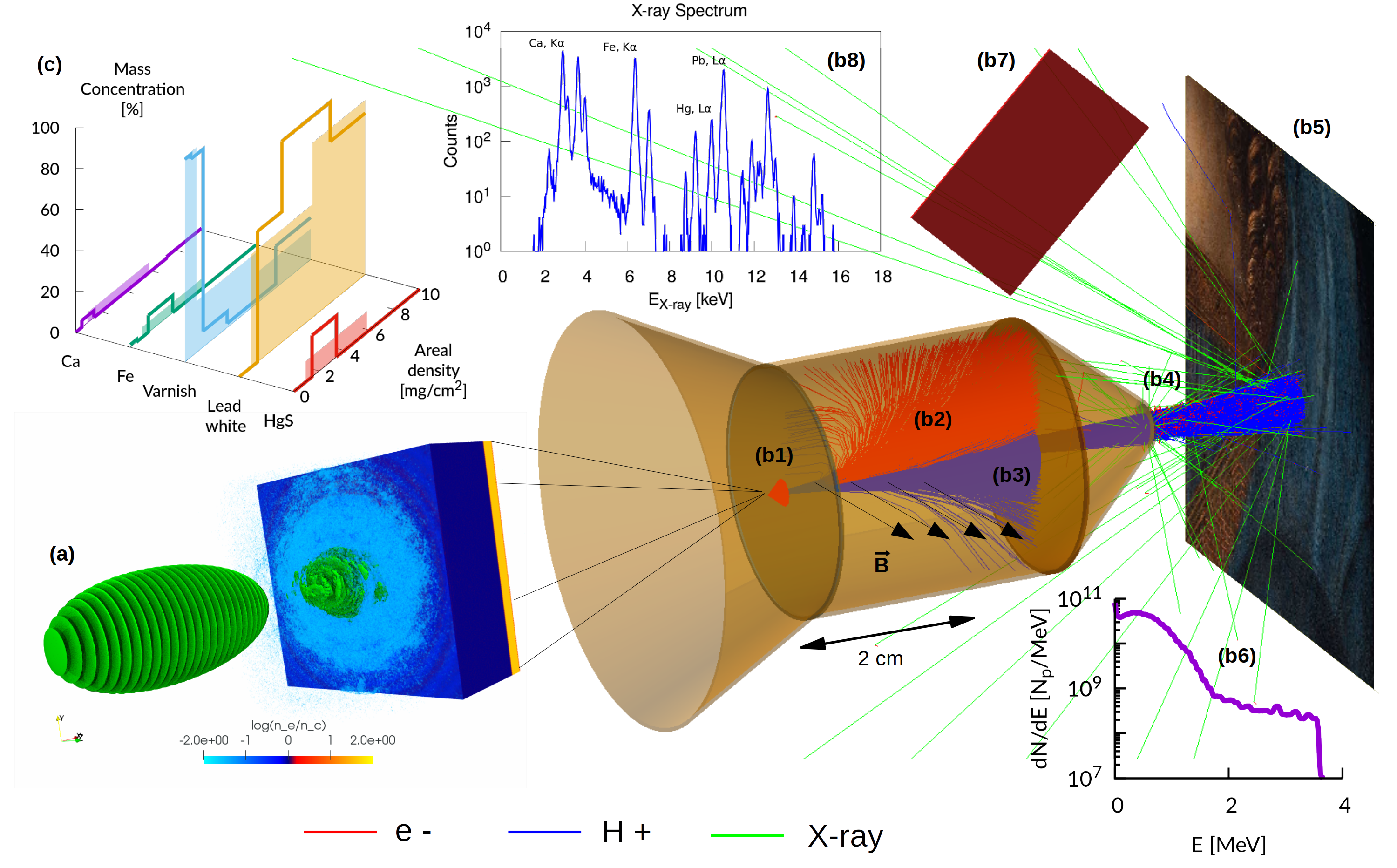}
\caption{Laser-driven in-air PIXE scheme. (a) Laser-driven ion source: a double-layer target irradiated with an ultraintense laser (snapshot of a PIC simulation); (b) Monte Carlo simulation setup (electrons shown in red, ions shown in blue and X-rays shown in green): (b1) first aperture slit; (b2) 2.5 cm long dipole magnet having a field strength of 0.15 T; (b3) second aperture slit; (b4) 1 $\mu$m thick exit window made of $Si_{3}N_{4}$ \cite{PalonenNIMB2016}; (b5) sample in air (2 cm from exit window); (b6) proton energy spectrum at sample surface; (b7) 2.6 $\times$ 2.68 $\textrm{cm}^2$ rectangular screen representing a PI-LCX:1300 CCD camera; (b8) Example of one of the proton energy spectra at sample surface; (c) retrieved elemental concentration profiles (solid lines) vs real ones (filled regions).}
\label{fig:setup}
\end{figure}

Materials characterization is of crucial importance for basic research as well as for a wide number of applications, ranging from virtually every field of advanced technology to cultural heritage purposes. Ion Beam Analysis (IBA)\cite{BirdIBA1990,VermaIBA2007} is an important family of modern analytical techniques to probe the composition and the surface structure of solid samples with MeV ion beams. IBA has found use in an impressive number of applications \cite{MalmqvistRadPhysChem2004}, including biomedical elemental analysis \cite{PetiboisAnalBioanalChem2008}, semiconductor industry \cite{KarydasASS2014}, cultural heritage studies\cite{MackovaNucPhysHerit2016}, forensic analysis\cite{WarrenForensics2002}, nuclear fusion research\cite{RubelVacuum2005}. The appeal of these techniques is due to their non-destructive nature and their unparalleled detection capabilities.\\
The layout of an IBA apparatus involves an ion source, a target chamber (where the sample is placed) and suitable diagnostics. The interaction of the ions with the sample generates a secondary emission, whose properties allow to retrieve  the structure and the elemental composition of the sample.\\
Despite the growing interest for IBA and its widespread adoption by industries and research institutes, in most cases the oldest accelerator technologies, mostly Van de Graaf and Tandem systems, are adopted. As a consequence, al the available IBA techniques have been developed considering only monochromatic, conventional accelerator-based ion beams. Conventional ion sources are invariably affected by a number of limitations: non-tunable energy, radioprotection issues \cite{Mitu_J_Radiol_2015}, high costs, non-portable size. Ion sources driven by superintense lasers represent a  very interesting alternative\cite{MacchiRevModPhys2013}. The investigation of  laser-driven particle sources is a thriving research area, as demonstrated by the growing worldwide investments in large-scale facilities\cite{DansonHPLSE2015}, such as the pan-European Extreme Light Infrastructure\cite{EditorialNatMat2015}. \\
The most relevant properties of laser-driven ion sources (maximum ion energy and ion number per laser shot) already exceed, in principle, the requirements for several IBA techniques. In addition, recent progress in exploiting non-conventional targets allows to achieve such beam parameters with reduced laser requirements, which may now realistically fall into the class of compact, table-top Terawatt systems \cite{BlancoNJP2017,GauthierAppPhysLett2017}. Moreover, laser-driven ion sources offer the possibility to easily and rapidly change the energy of the accelerated ions. This can be done simply by tuning the laser pulse energy, with concrete, major advantages in terms of flexibility. As far as radioprotection is concerned, laser-driven ion sources should be more manageable since the radiation fields originate from a small region surrounding the laser-target interaction point. Thus, laser-driven ion accelerators for IBA offer a concrete route for increased flexibility and more compact, cheaper systems with reduced radioprotection concerns, which could revolutionize the entire field. On the other hand, the peculiar nature of these sources, invariably characterized by a broad energy distribution\cite{MacchiRevModPhys2013}, requires to address novel, so-far unsolved, challenges to make laser-driven IBA a reality.\\
In this work we demonstrate the full potential of laser-driven ion sources for IBA, by means of a detailed theoretical-numerical investigation of a particularly interesting technique: Proton Induced X-ray/Gamma-ray Emission, PIXE/PIGE. Specifically, we address and solve the following points: i) extension of the theoretical models to interpret PIXE/PIGE data, in order to make them suitable for laser-driven ions ii) development of realistic numerical simulations of a laser-driven PIXE experiment to show the capability of retrieving material properties with the developed numerical methods iii) design of a proof-of-concept, transportable, laser-driven source for IBA (see figure \ref{fig:setup}), obtained combining a compact, TW-class laser\cite{TTmobileLaser,Quark45Laser} with a non-conventional targetry solution based on nanostructured double-layer targets \cite{NakamuraPoP2010,SgattoniPRE2012,PassoniPPCF2014,PrencipePPCF2016,PassoniPRAB2016,CialfiPRE2016}.
\section*{PIXE/PIGE theory with laser-driven ions}\label{sec:IBAwithTNSA}
\begin{figure}[ht]
\centering
\includegraphics[scale = 0.50]{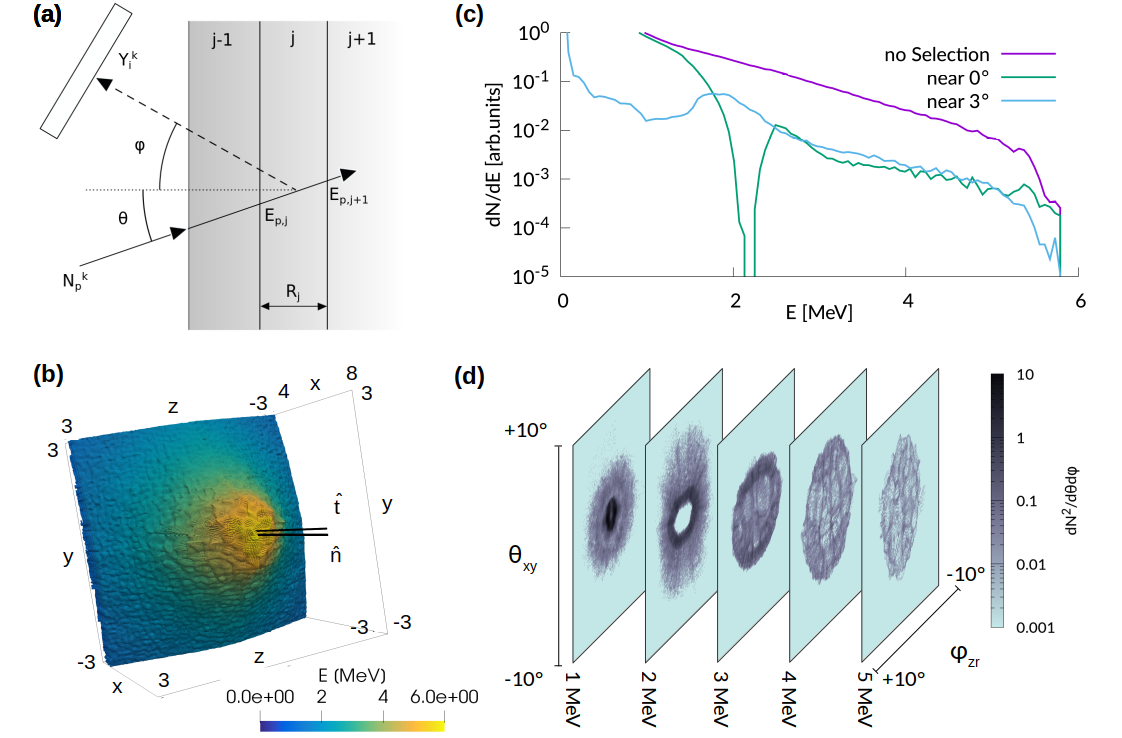}
\caption{\emph{Differential PIXE} scheme and properties of the laser-driven ion source. (a) Schematic representation of \emph{differential PIXE}. (b) Snapshot at 267 fs of the macro-protons for the PIC simulation performed with $\textrm{a}_{0} = 4.0$. Only the central region of the simulation box is shown and the macro-particles are colored according to their energy. The two black lines indicate the normal to the target surface ($\hat{n}$) and a slightly tilted direction ($\hat{t}$). (c) Proton energy spectra for all the macro-particles (purple), for the particles propagating along $\hat{n}$ (light blue) and $\hat{t}$ (green). (d) Angular and energy distribution of the protons for the PIC simulation performed with $\textrm{a}_{0} = 4.0$ (t = 267 fs). Each panel shows the angular distribution at the corresponding energy range.}
\label{fig:model}
\end{figure}

Laser-driven ions are invariably characterized by a broad-spectrum energy distribution\cite{MacchiRevModPhys2013}, whose shape depends on the specific acceleration mechanism. Since PIXE/PIGE formalisms have been developed only for monochromatic ion sources\cite{VermaIBA2007}, extending the theory for arbitrary energy distributions is a crucial step.
Consider first conventional PIXE analysis, performed on a thick homogeneous sample composed by $I$ elements of atomic weights $M_{i}$ and mass concentrations $W_{i}$, irradiated with a monoenergetic proton beam of initial energy $E_{0}$ and total number of protons $N_{p}$. The number of X-rays $Y_{i}$, associated to one characteristic line for the \textit{i}-th element, is \cite{VermaIBA2007, SmitNIMB2008}:
\begin{equation} \label{eq:Yield_omo_mono}
Y_{i} = N_{p} \frac{\Delta\Omega }{4\pi} \varepsilon_{i} \frac{N_{av}}{M_{i}}W_{i}\int_{E_{0}}^{E_{f}}\sigma_{i}(E)\omega_{i}e^{-\mu_{i}\int_{E_{0}}^{E'}\frac{dE'}{S(E')}\frac{\cos\theta}{\cos\phi}}\frac{dE}{S(E)} \equiv N_{p} \Gamma_{i}(E_{p})
\end{equation}
where $\Delta\Omega$ and $\varepsilon_{i}$  are the subtended solid angle and the efficiency of the detector, $N_{av}$ is the Avogadro's number, $E_{f}$ is the final proton energy after slowdown in matter, $\sigma_{i}(E)$ is the ionization cross section, $\omega_{i}$ is the fluorescence yield, $S(E)$ is the proton stopping power  in the sample, $\mu_{i}$ is the X-ray attenuation coefficient, $\theta$ is the proton impact angle and $\phi$ is the X-ray emission  angle. The integral takes into account the production of X-rays along the proton range, as well as their attenuation inside the material, and it is therefore computed over the projectile energy. \\
By neglecting the secondary fluorescence contribution, quantitative PIXE analysis can be performed by solving a system of $I$ equations like equation (\ref{eq:Yield_omo_mono}), where the X-ray yields $Y_{i}^{*}$ are measured experimentally and the unknowns are the mass concentrations $W_{i}$ of the elements. The system is non-linear because both the X-ray attenuation coefficients $\mu_{i}$ and the proton stopping power $S(E)$ depend upon the sample composition. Accordingly, the solution can be found by minimizing a parameter like $\chi^{2} = \sum_{i}[(Y_{i}-Y_{i}^{*})/\sqrt{Y_{i}^{*}}]^{2}$, making use of an iterative procedure. \\
If the sample is non-homogeneous, the goal of the analysis is to retrieve the concentration profiles of the element as a function of the thickness. In general, X-rays are emitted along the whole path of the protons inside the sample and the emission cannot be directly related to the deposited energy. In other words, it is not possible to simply assume that X-ray emission is representative of the sample composition at the Bragg peak. Therefore, a dedicated procedure must be followed as far as a non-homogeneous sample is of concern. In this case, \emph{Differential PIXE} analysis can be applied by ideally splitting the sample in a finite number $J$ of fictitious layers having mass thickness $\rho_{j}R_{j}$ and homogeneous composition $W_{i,j}$, as shown in figure \ref{fig:model}(a). Assuming to perform $K$ measurements at different proton energies $E_{p}^{k}$, the following system of equations can be  written \cite{SmitNIMB2004, SmitNIMB2008}:
\begin{equation} \label{eq:Yield_momo_diff}
Y_{i}^{k} = N_{p}^{k}\frac{\Delta\Omega }{4\pi} \varepsilon_{i} \frac{N_{av}}{M_{i}}\sum_{j=1}^{J-1} W_{i,j} e^{-\sum_{l=1}^{j-1}(\frac{\mu}{\rho})_{i,l} \frac{\rho_{l} R_{l}-\rho_{l} R_{l-1}}{\cos\phi}} \int_{E_{p,j+1}^{k}}^{E_{p,j}^{k}}\sigma_{i}(E)\omega_{i}e^{-\mu_{i,j}\int_{E_{p}}^{E'}\frac{dE'}{S_{j}(E')}\frac{\cos\theta}{\cos\phi}}\frac{dE}{S_{j}(E)}
\end{equation}
where $N_{p}^{k}$ is the number of protons with energy $E_{p}^{k}$, $(\frac{\mu}{\rho})_{i,j}$ and $S_{j}(E)$ are the X-ray attenuation coefficient and proton stopping power associated to the $j$-th layer, $E_{p,j}^{k}$ and $E_{p,j+1}^{k}$ are the proton energies at the boundaries of the  $j$-th layer (see figure \ref{fig:model}(a)). The system of $I \times K$ equations can be solved if $K \geq J$ (i.e. if a number of measurements greater than or equal to the number of fictitious layers is performed) by $\chi^{2}$ minimization. The solution is represented by the set of discrete values of $W_{i,j}$ and mass thicknesses $\rho_{j} R_{j}$ which best fit the experimental X-ray yields $Y_{i}^{k*}$.\\
We extended this formalism to account for a non-monoenergetic proton energy spectrum, represented by the function $f_{p}(E_{p}) = dN_{p}(E_{p})/dE_{p}$, which we assume to be different from zero in the energy interval from $E_{p,min}$ to $E_{p,max}$. The number of X-rays $dY_{i}$ generated by the incident protons with energy between $E_{p}$ and $E_{p} + dE_{p}$ is $dY_{i} = dN_{p} \Gamma_{i}=f_{p}\Gamma_{i}dE_{p}$. Therefore, integrating over the incident proton energy, equation (\ref{eq:Yield_omo_mono}) can be generalized as follows:
\begin{equation} \label{eq:Yield_omo_exp}
Y_{i} = \frac{\Delta\Omega }{4\pi} \varepsilon_{i} \frac{N_{av}}{M_{i}}W_{i}\int_{E_{p,min}}^{E_{p,max}}f_{p}(E_{p})\int_{E_{p}}^{0}\sigma_{i}(E)\omega_{i}e^{-\mu_{i}\int_{E_{p}}^{E'}\frac{dE'}{S(E')}\frac{\cos\theta}{\cos\phi}}\frac{dE}{S(E)}dE_{p}
\end{equation}
With the choice $f_{p} = N_{p}\delta (E_{p}-E_{0})$ (i.e. a monochromatic distribution) equation (\ref{eq:Yield_omo_exp}) reduces to equation (\ref{eq:Yield_omo_mono}).\\
Also the case of \emph{Differential PIXE} can be generalized for $K$ arbitrary energy distributions. Equation (\ref{eq:Yield_momo_diff}) becomes:
\begin{equation} \label{eq:Yield_nnomo_exp}
Y_{i}^{k} = \frac{\Delta\Omega }{4\pi} \varepsilon_{i} \frac{N_{av}}{M_{i}}\sum_{j=1}^{J-1} W_{i,j} e^{-\sum_{l=1}^{j-1}(\frac{\mu}{\rho})_{i,l} \frac{\rho_{l} R_{l}-\rho_{l} R_{l-1}}{\cos\phi}} \int_{E_{p,min}^{k}}^{E_{p,max}^{k}}f_{p}^{k}(E_{p})\int_{E_{p,j+1}^{k}}^{E_{p,j}^{k}}\sigma_{i}(E)\omega_{i}e^{-\mu_{i,j}\int_{E_{p}}^{E'}\frac{dE'}{S_{j}(E')}\frac{\cos\theta}{\cos\phi}}\frac{dE}{S_{j}(E)}dE_{p}
\end{equation}
Here, the quantities $f_{p}^{k}(E_{p})$, $E_{p,max}^{k}$ and $E_{p,min}^{k}$ refer to the $k$-th proton energy spectrum. Equation (\ref{eq:Yield_nnomo_exp}) implies that also in the case of broad spectrum sources it is necessary to use at least K different spectra to solve the system.\\
The formalism developed here for PIXE with arbitrary ion sources extends naturally also to PIGE\cite{MATEUS2005302}: the ionization cross section $\sigma_{i}(E)$ and the fluorescence yield $\omega_{i}$ must be replaced with the proper nuclear reaction cross section and isotopic abundance, while $\gamma$-ray attenuation (i.e. the exponential terms in all the equations) can be neglected. \\
If a laser-driven proton source is used, the shaping function $f_{p}(E_{p})$ can take different forms, depending on the specific ion acceleration process \cite{MacchiRevModPhys2013}. Target Normal Sheath Acceleration (TNSA) is arguably the most well understood laser-driven ion acceleration mechanism: it stands out for its robustness among other schemes \cite{BorghesiNIMA2016}, it has been used extensively as diagnostic tool in laser plasma interaction experiments \cite{Borghesi2002} and has been demonstrated in a repetitive regime \cite{NishiuchiPRAB2010}. TNSA can be outlined as follows: a micrometric solid foil is irradiated with an ultra intense laser pulse, which accelerates the electrons of the target towards the back side. The expansion of the hot electron cloud generates then an intense electric field, which drives ion acceleration. Accelerated ions (mostly protons from the hydrocarbon contaminants at the back side of the target) show an exponential energy distribution, with cut-off energies ranging from few MeVs for multi-TW class lasers\cite{BlancoNJP2017,GauthierAppPhysLett2017} up to  $\sim 100$ MeV \cite{WagnerPRL2016, HigginsonNatCom2018} for high energy large scale laser systems. The energy distribution can also depend significantly on the angular selection as well as on the subsequent propagation toward the sample region. Other non-TNSA acceleration schemes, such as Collisionless Shock Acceleration (CSA) \cite{haberberger2012collisionless} or Radiation Pressure Acceleration (RPA) \cite{HenigPRL2009, PalmerPRL2011, ScullionPRL2017} have been reported in the literature. These acceleration schemes can provide a narrower energy spectrum with respect to TNSA, though still much broader than conventional electrostatic accelerators, preventing  the use of the standard algorithms for PIXE with monochromatic sources. Therefore, CSA or RPA cannot be considered more suitable than TNSA for PIXE. The approach described here could nonetheless be used in principle also for CSA or RPA.\\
Quantitative laser-driven PIXE analysis, both considering homogeneous samples and \emph{Differential PIXE}, requires to implement an iterative algorithm to solve the system of equations (\ref{eq:Yield_omo_exp}) or (\ref{eq:Yield_nnomo_exp}) to retrieve the sample composition, analogous to those used for PIXE with monoenergetic protons. We will test the capability to retrieve material properties in unknown samples by performing complete simulations of laser-driven PIXE measurement, as detailed in the following. Reliable and complete experimental datasets including both energy spectra and angular distribution for several laser intensities are not available in the literature. Therefore, we will exploit both a simplified analytical modeling and a realistic numerical description of the laser-driven proton source.\\

\section*{Simulation of a laser-driven PIXE experiment} \label{sec:MC}
The goal of the present work is to assess the potential of laser-driven PIXE in realistic scenarios, i.e. demonstrate the possibility to retrieve the composition of complex, realistic samples from X-ray emission induced by a laser-driven proton source. To this aim, we consider a selection of artifacts directly relevant for cultural heritage studies (e.g. paintings, jewelry\ldots), a research field in which PIXE finds widespread use. The complexity of these artifacts, which naturally have a multi-elemental, multi-layer composition is an ideal testbed to demonstrate the general feasibility of laser-driven PIXE and \emph{Differential PIXE} for a wide range of applications. For the purpose outlined here, no useful experimental data are available in the literature, with only one, partial, exception. In \cite{BarberioSciRep2017} the possibility to obtain an X-ray spectrum irradiating ceramic and metallic samples with a laser-driven proton source was demonstrated, assessing also the absence of any damage to the irradiated samples. However, the full composition was not retrieved and the reported data are too simplified to be considered for the goals of this work.\\
Therefore, by means of Geant4 Monte Carlo simulations, we generate ``synthetic'' realistic X-ray spectra (i.e. the experimental X-ray yields $Y_{i}^{*}$ or $Y_{i}^{k*}$) in a variety of laser-driven PIXE scenarios (different artifacts and different laser-driven ion sources). For each scenario, the x-ray spectra are analyzed with an iterative code based on equations (\ref{eq:Yield_omo_exp}) and (\ref{eq:Yield_nnomo_exp}), as well as on an efficient minimization algorithm (see Methods section) designed to retrieve the sample compositions. These  compositions are then compared with the sample composition initially set in the Monte Carlo simulations. In the following we will refer to the elemental concentrations set in the Monte Carlo code as the \emph{real} concentrations and to the elemental concentrations retrieved by  the iterative code as the \emph{retrieved} concentrations. Specifically, we investigated a total of three scenarios (involving two different experimental set-ups).\\
For the first scenario we consider PIXE analysis of a homogeneous metallic sample in vacuum, performed with an idealized proton source. In this case, we choose  a simple analytical TNSA-like exponential for the energy spectrum of the proton source in the Monte Carlo: $f_{p}(E_{p})\sim\exp\left({-E_{p}/T_{p}}\right)$, characterized by a maximum cut-off energy $E_{p,max}$ and a parameter $T_{p}$ (related to the mean proton energy). We choose $E_{p,max} = 5 \textrm{ MeV}$ and $T_{p} = 0.7 \textrm{ MeV}$, which is realistic for a proton source driven by a 20 TW laser (see e.g. ref. \cite{GauthierAppPhysLett2017}, where TNSA is performed with a pure hydrogen target). The simple model adopted for the ion source allows also to perform a sensitivity analysis to address the effects of typical fluctuations of both $E_{p,max}$ and $T_{p}$ on a PIXE measurement. We consider a sample composed by iron, copper, zinc, tin and lead, which is representative of a Roman sword-scabbard \cite{SMIT200527}.\\
For the second scenario we consider \emph{Differential PIXE} analysis of a non-homogeneous metallic sample in vacuum performed with an idealized proton source. In this case we use 6 different analytical proton spectra with properties ($E_{p,max}$ and $T_{p}$) compatible with a laser system having a maximum power of 40 TW \cite{GauthierAppPhysLett2017} (highest $E_{p,max}$ equal to 6 MeV). In this scenario, the sample composition is inspired by the elemental concentration profiles of a medieval broach \cite{SMIT20082329}: three layers (a 1.2 $\mu$m superficial layer, a 1.3 $\mu$m intermediate layer and a thick substrate), with different concentrations of copper, zinc, silver, gold and mercury in each layer.\\
For the third scenario we consider \emph{differential PIXE} analysis of a non-homogeneous organic sample in air, performed with a realistic proton source obtained by means of Particle-In-Cell simulations. In particular, we model a realistic laser-driven ion source based on an advanced TNSA scheme, in which a double-layer target (a solid foil coated with a low density foam\cite{ZaniCarbon2013}) is used to increase laser-target coupling and thus to enhance the maximum energy and the number of accelerated ions\cite{NakamuraPoP2010,SgattoniPRE2012,PassoniPPCF2014,PrencipePPCF2016,PassoniPRAB2016}. In order to provide a realistic energy and angular distribution of the accelerated ions, we perform fully three-dimensional (3D) Particle-In-Cell (PIC) \cite{ArberPPCF2015} simulations of such laser-driven ion source. We consider a 800 nm thick solid foil coupled with a near-critical 5 $\mu$m plasma layer, irradiated with a 30 femtoseconds, ultra-intense (up to $\sim 5 \times 10^{19} \textrm{W/cm}^2$) laser pulse. These parameters are realistic for a tightly focused (waist of $3 \mu$m) 20 Terawatt Ti:Sapphire laser-system (e.g. they can be obtained scaling down the parameters reported in \cite{ZeilNatComm2012} for a 200 Terawatt laser, reducing the intensity while keeping the waist constant at 3 $\mu$m). Figures \ref{fig:model}(b-d) show the main properties of this laser-driven proton source. Figure \ref{fig:model}(d) shows a ring-like spatial distribution of the accelerated protons, which results into a hole in the energy spectrum at $0^{\circ}$ (fig. \ref{fig:model}(b)). Similar distributions have been observed frequently in the literature (see \cite{BeckerPPCF2018} and references therein). In order to test our technique on a broader distribution we decided to tilt our source by $\sim 0^{\circ}$. This means that the laser is still at normal incidence with respect to the target, but the laser axis (which is also the target normal axis) is tilted by $3^{\circ}$ with respect to the beam handler axis.
We use 6 different proton spectra whose properties are changed varing the intensity of the laser beam in the PIC simulation. We choose a complex multi-layer sample, whose stratigraphic structure is representative of an oil painting\cite{article_V, SmitNIMB2008}. The superficial layer has a thickness of 10 $\mu$m and it is composed by a protective organic varnish. The intermediate layer is 20 $\mu$m thick and contains a red pigment (HgS, the so-called \emph{Cinnabar}), a lead-based white pigment (the so-called \emph{lead-white}), some impurities (Ca and Fe) and the organic binder. The thick substrate (the so called \emph{imprimitura}) is composed only by the \emph{lead-white} and the organic binder. \\
The whole experimental apparatus assembled with Geant4 for the third  scenario (PIXE in air) is illustrated in figures \ref{fig:setup} (the set up for the first and second scenario neglects the presence of the air and exit window, being otherwise identical). Primary particles are generated at a punctual source placed in vacuum (figure \ref{fig:setup} (a)). In principle, in case of a double layer target, we should consider protons, electrons  and heavier ions. However we neglected heavier ions in the Monte Carlo simulations since in TNSA regime their number is usually orders of magnitude lower than for protons \cite{McGuffeyNJP2016, BinPRL2018}. Moreover their cut-off energy per nucleon is usually lower, further reducing the X-ray yield due to heavier ions. Laser-driven proton sources generate also a copious amount of energetic electrons \cite{MacchiRevModPhys2013}, which would produce an unwanted X-ray signal from the sample. Thus the apparatus in figure \ref{fig:setup} is designed to remove these energetic electrons. The particles are made to pass across an aperture slit (figure \ref{fig:setup} (b1)) and a dipole magnet (figure \ref{fig:setup} (b2)), which is a standard component to manipulate the laser-driven protons path \cite{DoriaAIPAdv2012, YogoAppPhysLett2009}. The magnetic field steers electrons and protons trajectories. The electrons are completely deflected and they are stopped by a second slit (figure \ref{fig:setup} (b3)). On the other hand, protons undergo a smaller deflection and they partially cross the second slit. Then, protons pass through an exit window (figure \ref{fig:setup} (b4)) and they travel across the air before reaching the sample (figure \ref{fig:setup} (b5)). The proton spectra at the sample surface are shown in figure \ref{fig:painting} (f). It is worth to point out that compact target chambers for laser-driven ion acceleration are already commercially available \cite{KaioSourceLab} and could be easily adapted to beam handlers. The X-ray detector has the features of a commercial Charged Coupled Device (CCD) (figure \ref{fig:setup} (b7)). The simulations account for the detection efficiency of the instrument. Under proper operating conditions (i.e. number of incident photons lower than the number of pixels), CCDs are able to perform single shot X-ray spectrum measurements\cite{HongChinPhysB2017}. This feature makes CCDs an interesting choice in order to record the X-ray spectra emitted by the sample during irradiation with laser-driven protons (figure \ref{fig:setup} (b8)).\\

\section*{Results} \label{results}
We discuss the results of the three scenarios described in the previous section: PIXE analysis of a metallic homogeneous sample, \emph{Differential PIXE} analysis of a metallic non-homogeneous sample and \emph{Differential PIXE} analysis of a painting. We show that in all these cases the iterative algorithm is able to retrieve elemental concentration profiles in very good agreement with the real ones. All the details concerning the $\chi^{2}$ minimization procedure are reported in the Methods section. \\
Typical PIXE measurements on delicate artifacts are performed with beam currents of the order of tens of pA for 100s seconds \cite{MackovaNucPhysHerit2016}. This means $\sim 10^{10}$ protons on sample. Considering protons with energy greater than 0.2 MeV at the source, the efficiency of our beam-handling setup is $\sim 10 \%$ in the most challenging case (i.e. the in-air laser-driven PIXE analysis and $a_0 = 2$). Therefore, we estimate that the laser-driven ion source should produce $\sim 10^{11}$ protons with energy greater than $\sim 0.2$ MeV at the source point. This can be achieved with $\sim 10$s of shots with a sub-100 TW class laser \cite{NishiuchiPRAB2010} (which can be operated at $\sim 1$ Hz). In this case the number of generated X-rays per shot would be low enough to avoid saturation of the CCD. It is also worth to point out that single-shot PIXE measurements can be performed with $\sim 1$ PW class laser, provided that a passive X-ray detector is used \cite{BarberioSciRep2017}.\\
As shown in figures \ref{fig:metal}(g) and \ref{fig:painting}(g), in all the cases considered here the probed region on the samples is $\sim 1 \textrm{mm}^{2}$ (the probed region is not shown for the homogeneous sample analysis, however it is identical to that of fig. \ref{fig:metal}(g)). This is perfectly adequate for most applications \cite{ENGUITA200453}. \\

\subsection*{PIXE analysis of a metallic homogeneous sample}
\begin{figure}[ht]
\centering
\includegraphics[scale=0.13]{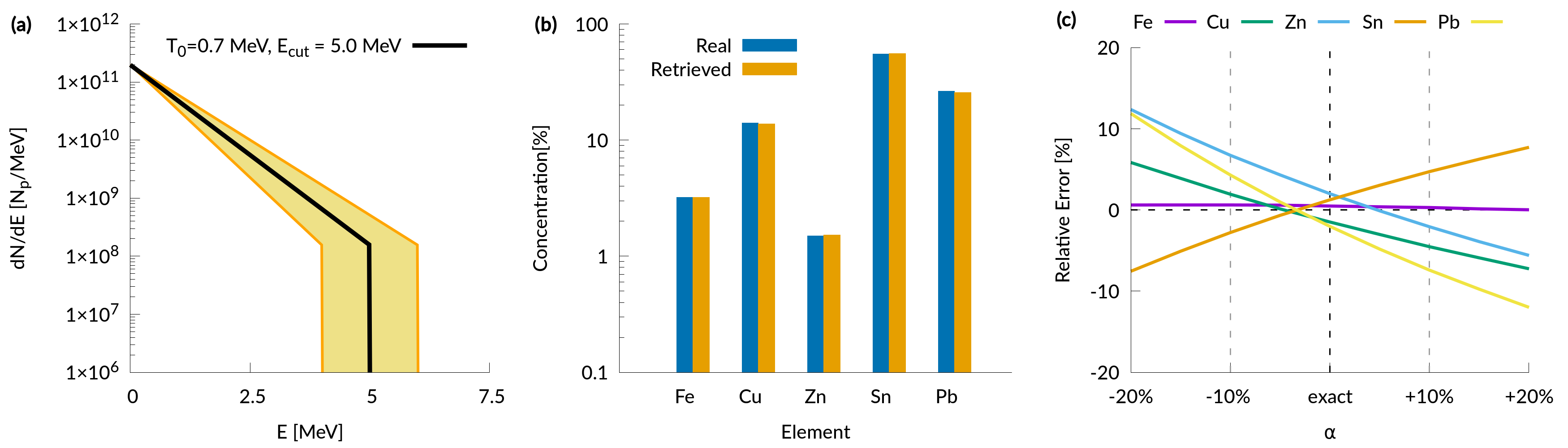}
\caption{Homogeneous sample results. (a) proton energy spectrum used in the Monte Carlo simulation (black curve) and energy spectra used to retrieve elemental concentrations for the sensitivity analysis (filled region); (b) Comparison between the retrieved elemental concentrations and the real ones; (c) relative error of the retrieved concentrations as a function of the fluctuation parameter $\alpha$.}
\label{fig:homo}
\end{figure}

In the first scenario, an idealized proton source having a pure exponential energy distribution (see figure \ref{fig:homo} (a)) and a few degrees divergence is considered. Figure \ref{fig:homo} (b) shows that the agreement between the real concentrations and the retrieved ones is very good. For all the elements the discrepancy is less than 1 $\%$, which is comparable to the precision expected for traditional PIXE. However this excellent agreement has been obtained assuming a perfect knowledge of the proton energy spectrum (i.e. the black curve in fig. \ref{fig:homo} (a)). On the other hand, it is well known that laser-driven proton sources are less stable than conventional particle accelerator sources, showing shot to shot fluctuations of the energy spectrum. These fluctuations might in principle hinder the applicability of laser-driven PIXE, thus it is imperative to asses their effect on the retrieved concentrations. This can be done straightforwardly for the simple scenario considered here. Keeping the X-ray yields obtained from the Monte Carlo simulations, we applied the iterative code considering a proton energy spectrum described by $f_{p}(E_{p},\alpha)\sim\exp\left({-E_{p}/(\alpha T_{MC})}\right) \Theta(E_p - \alpha E_{MC})$ (i.e. $\alpha = 1$ means that the iterative code uses a proton energy spectrum identical to that of the Monte Carlo simulation). $\alpha$ was varied between 0.8 and 1.2 in order to test the effect of fluctuations up to $\pm 20 \%$, as shown in figure \ref{fig:homo}(a). Figure \ref{fig:homo}(c) reports the ratio between the concentrations obtained with a given $\alpha$ and those obtained with $\alpha = 1$. The graph shows a mild dependence of the relative error on the fluctuations of the proton energy spectrum: even for large ($\pm 20 \%$) fluctuations, the relative error remains approximatively between $\pm 10 \%$.
\subsection*{\emph{Differential PIXE} analysis of a metallic non-homogeneous sample and of a  painting}
\begin{figure}[ht]
\centering
\includegraphics[width = 0.85\textwidth]{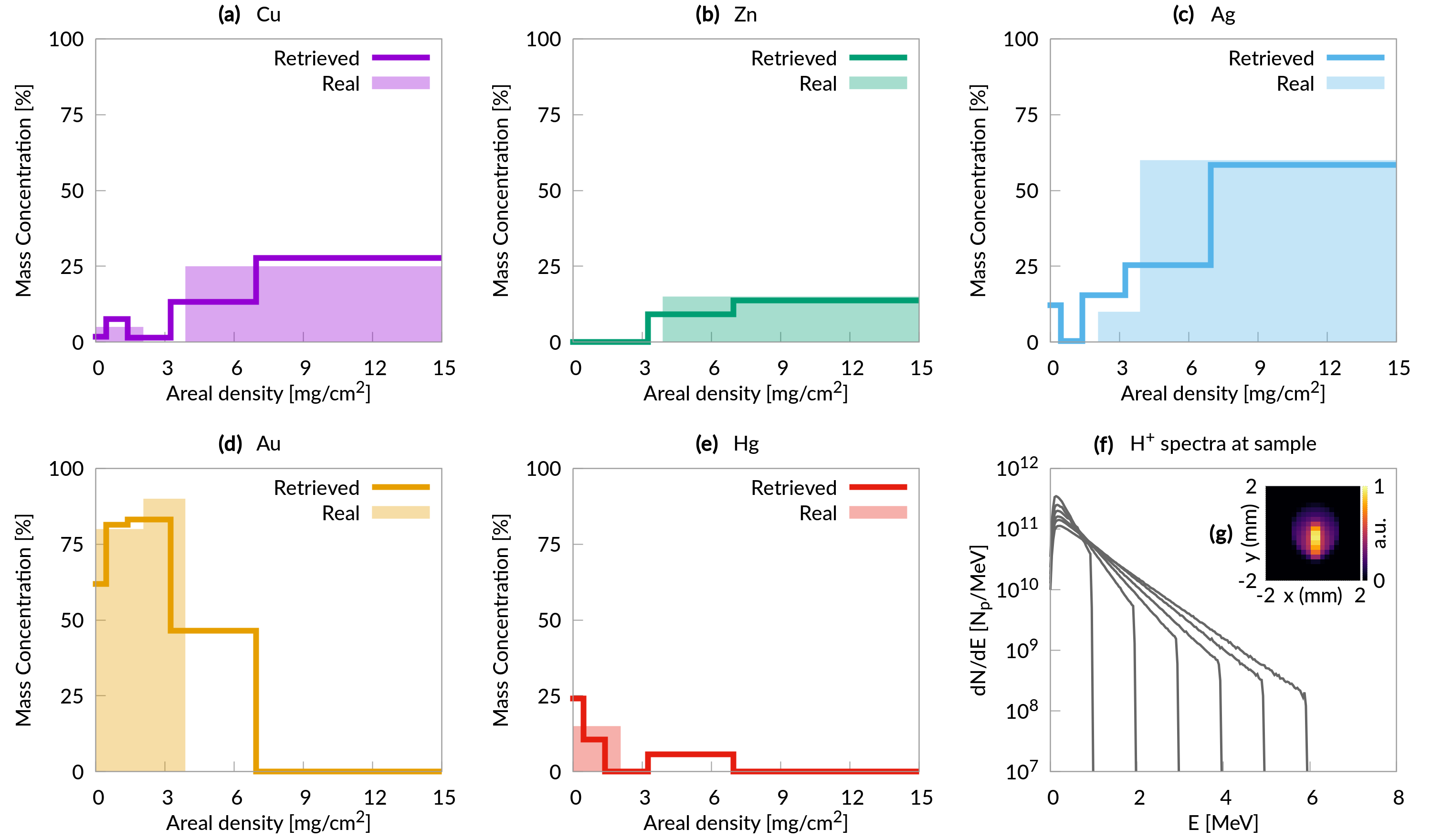}
\caption{Non-homogeneous metallic sample results. (a-e) Comparison between the retrieved elemental concentration profiles and the real ones; (f) proton energy spectra at the sample surface; (g) proton fluence at the sample surface.}
\label{fig:metal}
\end{figure}

\begin{figure}[ht]
\centering
\includegraphics[width = 0.85\textwidth]{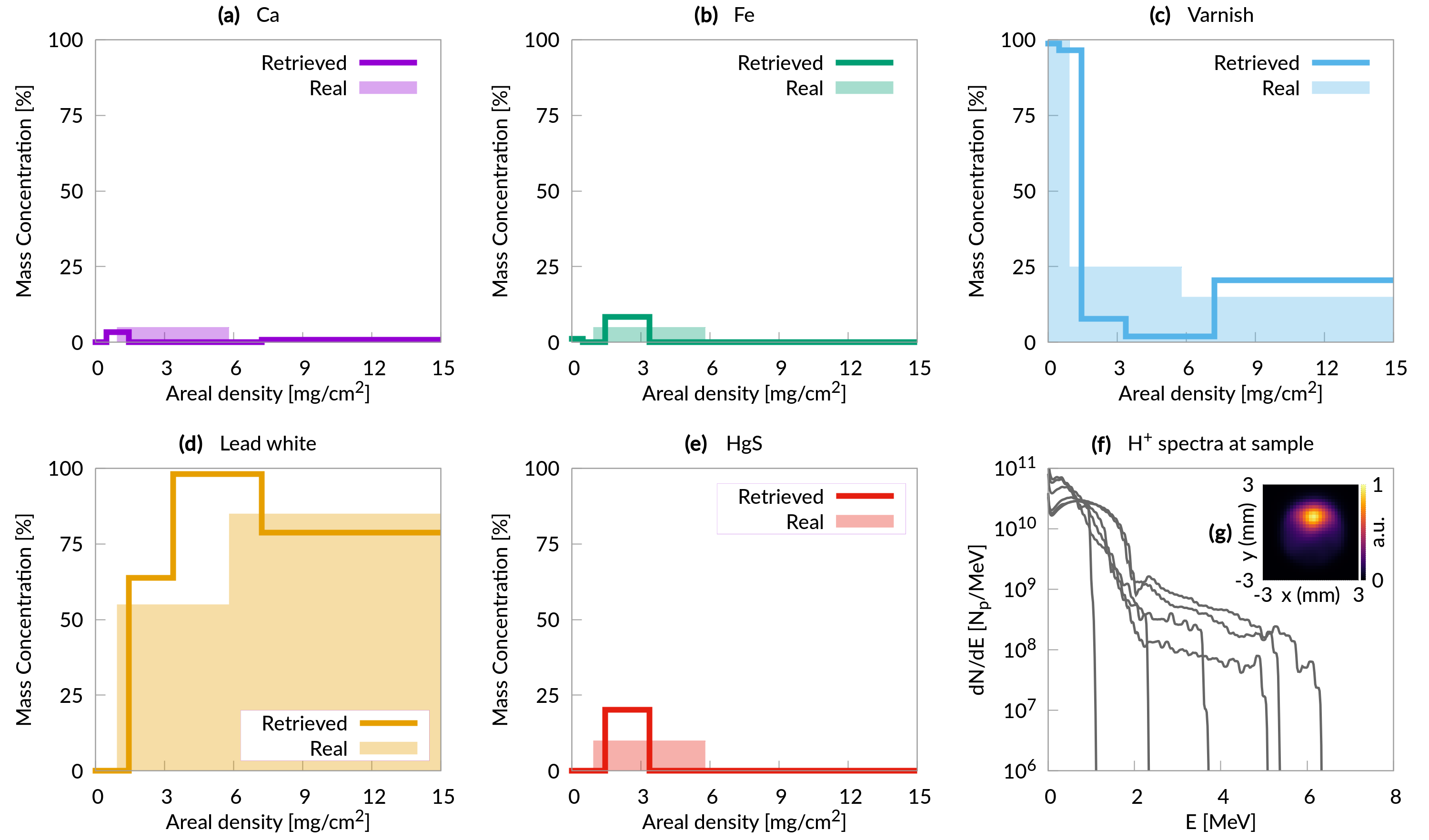}
\caption{Painting results. (a-e) Comparison between the retrieved elemental concentration profiles and the real ones; (f) proton energy spectra at the sample surface; (g) proton fluence at the sample surface.}
\label{fig:painting}
\end{figure}

Also in the second scenario, an idealized proton source was considered: six different exponential distributions with cut-off energies between 1 and 6 MeV (figure \ref{fig:metal}(f)). 
The concentration profiles (figures \ref{fig:metal}(a-b)), reconstructed from the X-ray yields, show a good agreement with the original ones. 
In the third scenario (the painting) the output of six PIC simulations is used as a proton source in the Monte Carlo code (figure \ref{fig:painting}(f) shows the energy spectra at the sample surface). Concerning the analysis discussed here, the main difference with respect to the metallic case is that now the recorded X-ray peaks are representative of complex compounds rather than individual elements. 
The analysis clearly identifies the superficial protective varnish, the intermediate region containing the pigment, and the substrate. The concentration profiles of the elements are in satisfactory agreement with the real ones. Therefore, the system considered in this scenario (based on a 20 TW system and on the use of double layer targets) represents a realistic design for a laser-driven PIXE apparatus.
\section*{Conclusions} \label{theend}
Laser-driven ion sources have been a thriving research topic in the last 18 years, thanks to their many foreseen applications in a number of scientific, technological and societal fields. While for most of such applications major challenges still need to be tackled \cite{FritzlerAppPhysLett2003, LinzPRAB2016}, the results presented in this work {for laser-driven PIXE clearly demonstrate the feasibility of a compact, versatile, cost-saving and portable laser-driven apparatus for IBA. This can be obtained combining a thorough theoretical description of the process with the most recent advanced laser-driven ion acceleration schemes and state-of-the-art laser systems. One can easily imagine further exciting developments, for example the possibility to exploit the same ion source for the production of neutrons with unique properties \cite{brenner:ppcf:2015}, paving the way for a true revolution in the field of hadron-based advanced material characterization techniques.

\section*{Methods} \label{instruments}
We developed an iterative code, based on  $\chi^{2}$ minimization algorithm, which solves the system of equations (\ref{eq:Yield_omo_exp}) or (\ref{eq:Yield_nnomo_exp}). We used this code to analyze the X-ray spectra obtained from Monte Carlo simulations of laser-driven PIXE experiments. For the third scenario, we performed Particle-In-Cell simulations to generate a realistic proton source.

\subsection*{$\chi^{2}$ minimization algorithm}
The minimization algorithm used to retrieve the elemental composition of the samples relies on the open source \emph{dlib C++ library} \cite{king2009dlib}, in particular on the \emph{BOBYQA} algorithm. All the physical parameters used in the algorithm are taken from standard libraries. For the ionization cross sections we used the Energy-Loss Coulomb-Repulsion Perturbed-Stationary-State Relativistic (ECPSSR) theory \cite{Brandt1979, Brandt1981}. For fluorescence yields we used values tabulated in the Evaluated Atomic Data Library (EADL) \cite{Perkins:236347, cullen1992program}. For X-ray mass absorption coefficients we used the online interface of the XCOM code  \cite{xcom_2010}. For the proton stopping powers we used the Stopping and Range of Ions in Matter (SRIM) code \cite{ZIEGLER20101818}. \\
For the homogeneous sample, as initial guess, the elements are assumed to be equally concentrated.\\
The \emph{differential PIXE} analysis demands an initial guess also on the mass thickness of the fictitious layers. This choice should take into account the expected penetration depth of the protons and the expected scalelength of the inhomogeneities of the sample. We chose $[0.65,1.3,2.6,5.2,\infty]$ $\textrm{mg}/\textrm{cm}^2$ both for the metal sample and for the painting. However, we introduced a free-parameter to readjust the mass thicknesses of the fictitious layers (e.g. $[0.65\xi,1.3\xi,2.6\xi,5.2\xi,\infty]$ $\textrm{mg}/\textrm{cm}^2$ with $\xi$ constrained between 0.5 and 1.5). The fact that the analysis converges to good results for both the metallic sample and the painting, despite their very different structure, confirms the generality of this approach. As far as the initial guess on the  concentrations is of concern, we followed a procedure analogous to that described in \cite{SmitNIMB2008}. \\
For all the cases considered in this work, the $\chi^{2}$ parameter has the general form:
\begin{equation} \label{Chi_q_metodi}
\chi^{2} = \sum_{i,k}\Big(\frac{\tilde{Y}_{i}^{k}-\tilde{Y}_{i}^{k,*}}{\sqrt{\tilde{Y}_{i}^{k,*}}}\Big)^{2}
\end{equation}
where the terms $\tilde{Y}_{i}^{k}$ and $\tilde{Y}_{i}^{k,*}$ are explained below. The mass concentrations $W_{i}$ (where $i$ is the element index) are bounded between 0 and 1. In the case of the homogeneous sample analysis, only one measurement is necessary (i.e. the sum over $k$ disappears). $\tilde{Y}_{i}$ and $\tilde{Y}_{i}^{*}$ in  equation (\ref{Chi_q_metodi}) are, respectively, the normalized calculated yield and the normalized experimental yield (i.e. $\tilde{Y}_{i} = Y_{i}/\sum_{i}Y_{i}$ and $\tilde{Y}_{i}^{*} = Y_{i}^{*}/\sum_{i}Y_{i}^{*}$). This formulation allows to perform the $\chi^{2}$ minimization without knowledge of the number of incident protons, which is particularly interesting from the experimental point of view. \\
As far as the \emph{differential PIXE} analysis is concerned, $\chi^{2}$ is a function of the free-minimizing parameter $\xi$ and of the elemental concentrations $W_{i,j}$ of all the fictitious layers, which means 26 variables in the cases of interest here.
 Since the direct minimization of such function is particularly challenging we subdivided the procedure in multiple steps. We minimize $\chi^{2}$ one element at a time (together with $\xi$) for all the elements and we repeat this procedure until the convergence is reached. Moreover, we do not used the normalized X-ray yields, but $\tilde{Y}_{i}^{k}$ and $\tilde{Y}_{i}^{k,*}$ coincide with calculated $Y_{i}^{k}$ and experimental $Y_{i}^{k,*}$ ones.
Finally, in the case of the painting, the presence of a compound which does not provide detectable X-ray yields (i.e. the organic matrix) demands to impose the additional constrain  $\sum_{i}W_{i,j} = 1$. Thus, we added a term $\lambda \sum_{i}{(1-\sum_{i}W_{i,j})}^2$ to equation (\ref{Chi_q_metodi}), where $\lambda$ is a large number.

\subsection*{Particle In Cell (PIC) simulation}
3D Particle-In-Cell simulations were performed with the open-source code \emph{piccante}\cite{arxiv_piccante}. We used a computational box of $120 \lambda \times 80 \lambda \times 80 \lambda$ with a spatial resolution of 40 points per $\lambda$ along $\hat{x}$ and 15 points per $\lambda$ along $\hat{y}$ and $\hat{z}$, where $\lambda = 800$nm is the laser wavelength. Time resolution is set at 98\% of the Courant Limit. The P-polarized laser pulse had a $\cos^2$ temporal profile (intensity FWHM of 30 fs) and a transverse Gaussian profile (i.e. $E \sim E_{0} \exp\left({- r^{2}/w^{2}}\right)$ where $E$ is the electric field, $E_{0}$ is the maximum electric field, $r$ is the radius and $w$ is the waist of 3 $\mu$m). The normalized peak laser intensity $a_0$ was set to 2, 2.5, 3,  3.5, 4, 4.5 (corresponding to a focused intensity between $8.7 \cdot 10^{18} ~ \textrm{W/cm}^{2}$ and $4.4 \cdot 10^{19} ~ \textrm{W/cm}^{2}$). The incidence angle was $0^\circ$. The target consisted in a 5 $\mu$m low-density layer with a density equal to 1 $n_c$ (sampled with 4 macro-electrons and 1 macro-ion with Z/A = 0.5 per cell), an overdense 800 nm thick foil with a density of 40 $n_c$ (sampled with 40 macro-electrons and 8 macro-ions with Z/A = 0.5 per cell) and a 80 nm thick contaminant layer with a density of 5 $n_c$.
In order to simulate a fully ionized hydrocarbon contaminant layer, the 5 $n_c$ ion charge was partitioned as follows: 4.285 $n_c$ for a species with Z/A = 0.5 and 0.715 for a species with Z/A = 1. 64 macro-electrons per cell and 125 macro-ions per species per cell where used for the contaminant layer. The electron population was initialized with a small ($\sim$ eV) temperature. Periodic boundary conditions are used for both EM field and particles, however the box is large enough to avoid unphysical effects on the ion acceleration process. The duration of the simulations was 100 $\lambda/c$ = 267 fs.
A total of 6 PIC simulations was performed, one for each value for the normalized intensity $a_0$. While in principle 3D PIC simulations can provide an absolute number for the total accelerated charge, this number depends on the choice of some parameters (namely the thickness and density of the contaminant layer) which might differ from those of a real experiment. For this reason in figure \ref{fig:model}(c) we reported the energy spectra using arbitrary units.\\
In order to use PIC results for the proton source in the Monte Carlo code, we calculated the distribution function $d^{3}N/dp_{x}dp_{y}dp_{z}$ from the macro-proton phase space ($p_{x}$, $p_{y}$ and $p_{z}$ are the Cartesian components of the momentum).

\subsection*{Geant4 Monte Carlo simulation toolkit}
Geant4 (Geometry and Tracking) \cite{AgostinelliNIMA2003} is an abstract C++ base classes' toolkit dedicated to the Monte Carlo simulation of particles' transport through matter. It allows to implement the particle transport physics and the creation of secondary particles, covering a wide range of energies (from eV to TeV). Geant4 allows to simulate PIXE\cite{PiaIEEE2009} reliably, as validated by several recent works\cite{FrancisNIMB2013,AhmadNIMB2013,IncertiNIMB2015}. \\
Our Monte Carlo simulations take into account protons, electrons, positrons, and photons. The secondary charged particles are tracked considering Bremsstrahlung, ionization and multiple scattering, while photoelectric effect, Compton scattering and pair production are activated for photons. 
The production cut is set equal to 0.5 $\mu$m for secondary particles in order to avoid infrared divergence. As far as these physical processes are concerned, we used the EmStandardPhysics\_option3 module recommended by the Geant4 documentation for high accuracy with electrons and ions tracking. To properly model the proton ionization cross sections, the Energy-Loss Coulomb-Repulsion Perturbation-Stationary-State Relativistic Theory (ECPSSR) \cite{BrandtPRA1981} is selected. \\
In all the considered cases, the primary proton source is punctual and the primary particles can be electrons or protons.\\
Electrons are generated extracting their energy and angular divergence from an exponential and a uniform distributions respectively. Their energy spectrum was fitted from the PIC simulation with $a_0 = 4.5$. This is the worst case scenario since the highest energy electrons are generated.\\
As far as protons are concerned, for the metallic samples analysis a similar approach has been followed: the energy is extracted from an exponential distribution and a uniform angular distribution is assumed.\\
In the last case (i.e. the painting analysis), the proton source is modeled with the actual PIC simulations results. A PIC simulation provides a distribution of momenta. The primary proton momentum components are extracted with the Inverse Transform Sampling method from the PIC distribution and they are used to define the initial energy and propagation direction. To this aim we created a specific C++ class.\\
Several Geant4 simulations with different seeds of the Random Number Generator (RNG) were run in parallel and the final results were aggregated. A total of few $10^{10}$ - $10^{11}$ primary protons were simulated for each case. We relied on the “HepJamesRandom” generator, which implements the Marsaglia-Zaman RANMAR algorithm. This RNG has the essential property of providing a large number ($\sim 8\cdot 10^{8}$) of independent and very long sequences of pseudo-random numbers \cite{JamesCPC1990}.\\
It worth to point out that, since the Monte Carlo code allows to simulate one particle at a time, possible collective effects are neglected. This should not affect our results since collective effects have been reported to be negligible in this regime \cite{sciscio2018design}.

\subsection*{Data availability}
The data supporting the findings of this work are available from the corresponding authors on request.

\section*{Author contribution statement}
M.P. conceived the project and supervised all the activities. F.M. wrote the iterative code
with the assistance of L.F, developed the theoretical model and performed the Monte Carlo simulations. 
L.F. performed the Particle-In-Cell simulations. All the authors contributed equally to the 
preparation of the manuscript.
 
\section*{Acknowledgments}
This project has received funding from the European Research Council (ERC) under the European Union’s Horizon 2020 research and innovation programme (ENSURE grant agreement No 647554). We also acknowledge LISA and Iscra access schemes to MARCONI HPC machine at CINECA(Italy) via the projects  LAST, LIRF and EneDaG.

\section*{Competing interests}
The authors declare no competing interests.

\bibliography{biblio}

\end{document}